\documentclass[a4paper,fleqn,usenatbib]{mnras}

\usepackage{newtxtext,newtxmath}

\usepackage[T1]{fontenc}
\usepackage{ae,aecompl}

\usepackage{graphicx}	
\usepackage{amsmath}	
\usepackage{amssymb}	
\usepackage{siunitx}
\usepackage{makecell}
\usepackage{multirow}
\usepackage{color}
\usepackage[bottom]{footmisc}

\title[Machine learning structure formation]{Machine learning cosmological structure formation}

\author[L. Lucie-Smith et al.]{
Luisa Lucie-Smith,$^{1}$\thanks{E-mail: luisa.lucie-smith.15@ucl.ac.uk}
Hiranya V. Peiris,$^{1,2}$
Andrew Pontzen,$^{1}$
Michelle Lochner$^{1,3,4}$
\\
$^{1}$Department of Physics \& Astronomy, University College London, Gower Street, London WC1E 6BT, UK\\
$^{2}$The Oskar Klein Centre for Cosmoparticle Physics, Stockholm University, AlbaNova, Stockholm, SE-106 91, Sweden\\
$^{3}$African Institute for Mathematical Sciences, 6 Melrose Road, Muizenberg, 7945, South Africa\\
$^{4}$SKA SA, The Park, Park Road, Cape Town 7405, South Africa
}

\date{Accepted XXX. Received YYY; in original form ZZZ}

\pubyear{2017}

\begin{document}
\label{firstpage}
\pagerange{\pageref{firstpage}--\pageref{lastpage}}
\maketitle

\begin{abstract}
We train a machine learning algorithm to learn cosmological structure formation from N-body simulations. The algorithm infers the relationship between the initial conditions and the final dark matter haloes, without the need to introduce approximate halo collapse models. We gain insights into the physics driving halo formation by evaluating the predictive performance of the algorithm when provided with different types of information about the local environment around dark matter particles. The algorithm learns to predict whether or not dark matter particles will end up in haloes of a given mass range, based on spherical overdensities. We show that the resulting predictions match those of spherical collapse approximations such as extended Press-Schechter theory. Additional information on the shape of the local gravitational potential is not able to improve halo collapse predictions; the linear density field contains sufficient information for the algorithm to also reproduce ellipsoidal collapse predictions based on the Sheth-Tormen model. We investigate the algorithm's performance in terms of halo mass and radial position and perform blind analyses on independent initial conditions realisations to demonstrate the generality of our results. 
\end{abstract}




\section{Introduction}
Dark matter haloes are the fundamental building blocks of cosmic large-scale structure, and galaxies form by condensing in their cores. Understanding the structure, evolution and formation of dark matter haloes is an essential step towards understanding how galaxies form and ultimately, to test cosmological models.
However, this is a difficult problem due to the highly non-linear nature of the haloes' dynamics. Dark matter haloes originate from random perturbations seeded in the early Universe and grow via mass accretion and mergers with smaller structures throughout their assembly history. N-body simulations provide the only practical tool to compute non-linear gravitational effects starting from an initial random field \citep[{e.g.}][]{gadget, gadget2, Sim-stateofart}.

Analytic approximations of structure formation yield useful physical interpretations of these detailed numerical studies. Generally, analytic techniques assume dark matter collapse occurs once the smoothed linear density contrast exceeds a threshold value. Combined with excursion set theory, this ansatz provides a tool to analytically predict the final halo mass of an initially overdense region. This can be used to infer useful quantitites such as the abundance of dark matter haloes in the Universe, or the halo mass function, based on properties of a Gaussian random field alone \citep{PS, Bond, Bond&Myers}. The halo mass function is the quantity most often used to assess the accuracy of different analytic frameworks against numerical simulations. The original form of the halo mass function proposed by \citet{PS}, although qualitatively correct, is known to underestimate the abundance of the most massive haloes, and overestimate the abundance of the less massive ones. The need for precision mass functions led to modifications of the original halo mass function in the form of parametric functions calibrated with cosmological simulations \citep{Jenkins, Reed, Tinker}. Pure analytic extensions of the excursion set ansatz have also been constructed which yield better agreement with numerical simulations \citep{Sheth, Maggiore, Paranjape, Fahari, Porciani}. Given these successful predictions, the excursion set description has become an accepted physical interpretation of the process of structure formation itself.

We present a machine learning approach to learn cosmological structure formation directly from N-body simulations. The machine learning algorithm is trained to learn the relationship between the initial conditions and final halo population that results from non-linear evolution. Using the resulting initial conditions-to-haloes mapping, we aim to provide new physical insights into the process of dark matter halo formation, and compare with existing interpretations gained from widely investigated analytic frameworks. In contrast to existing analytic theories, our approach does not require prior assumptions about the physical process of halo collapse; the haloes' non-linear dynamics is learnt directly from N-body simulations rather than approximated by an excursion set model in the presence of a collapse threshold.

We provide the machine learning algorithm with a set of informative properties about the dark matter particles extracted from the initial conditions. Machine learning algorithms are sufficiently flexible to include a wide range of initial conditions properties which may contain relevant information about halo formation, without changing the training process of the algorithm. We choose these properties to be aspects of the initial density field in the local surroundings of the dark matter particles' initial position. By quantifying their impact on the learning accuracy of the algorithm, we can investigate which aspects of the early universe density field contain relevant information on the formation of dark matter haloes. The trained initial conditions-to-haloes mapping can then also be used to predict the mapping for new initial conditions, without the need to run a further simulation. 

The highly non-linear nature of dark matter evolution makes it a problem well-suited to machine learning. Machine learning is a highly efficient and powerful tool to learn relationships which are too complex for standard statistical techniques \citep{witten2016data}. In the context of structure formation, machine learning techniques have also been shown to be effective, for example, in learning the relationship between dark and baryonic matter from semi-analytic models \citep{ML-cosmology, Agarwal2017, Nadler2017}. 

We choose \emph{random forests} \citep{breiman1984classification, Leo}, a popular algorithm which has been shown to outperform other classifiers in many problems \citep{Niculescu-MizilCaruana, Caruana, Douglas, Lochner}. Random forests also lend themselves to physical interpretation, as they provide measures that allows the user to infer which of the inputs are predominantly responsible for the learning outcomes of the algorithm. Random forests are ensembles of decision trees, each following a set of simple decision rules to predict the class of a sample \citep{ballbrunner}. The prediction of the random forest is given by the average of the probabilistic predictions of the individual trees, where the variance of the forest predictions is greatly reduced compared to that of a single tree. 

To apply this approach, we must turn the process of dark matter evolution into a supervised classification problem. We chose to focus on the simplest case of a binary classification task to illustrate the approach and allow for a cleaner understanding of the physics behind the learning process of the algorithm. We distinguish between dark matter particles which end up in haloes of mass above a threshold, and those which belong either to lower mass haloes or to no halo at all. This defines two classes; the former set of particles belongs to the \emph{IN haloes} class while the latter forms the \textit{OUT haloes} class. The machine learning algorithm is trained to predict whether the dark matter particles in the initial conditions will end up in IN class haloes or in the OUT class at $z=0$. The training is performed on an existing N-body simulation where we already know the associated halo for each particle (if any). 

The predictive accuracy of the algorithm crucially depends on the choice of features extracted from the initial conditions and used as input to the machine learning algorithm. We first train the random forest with the initial linear density field as features and subsequently add information on the tidal shear field. We are able to quantify the physical relevance of such properties in the halo collapse process, based on their respective impact on the classification performance of the random forest. Our results demonstrate the utility of machine learning in gaining insights into the physics of structure formation, as well as providing a fast and efficient classification tool.

The paper is organised as follows. We present an overview of the classification pipeline and describe how we extract features from the linear density field and train the machine learning algorithm in Sec. \ref{sec:method}. In Sec. \ref{sec:denclass} we interpret the classification output and present our results in Sec. \ref{sec:den_results}. We then extend the feature set to include the tidal shear field in Sec. \ref{sec:shear} and discuss the resulting implications. We study the algorithm's performance as a function of halo properties in Sec. \ref{sec:part_properties}. We perform two blind tests of our pipeline on independent simulations in Sec. \ref{sec:blind}, demonstrating the generality of our results, and finally conclude in Sec. \ref{sec:conclusions}.

\section{Method}
\label{sec:method}
We trained and tested the random forest with an existing dark-matter-only simulation produced with \texttt{P-GADGET-3} \citep{gadget2, gadget} and a WMAP5 $\Lambda$CDM cosmological model \citep{WMAP}; $\Omega_{\Lambda} = 0.721$, $\Omega_{\mathrm{m}} = 0.279$, $\Omega_{\mathrm{b}} = 0.045$, $\sigma_{8} = 0.817$, $h = 0.701$, $n_s = 0.96$. The comoving softening length of the simulation is $\epsilon = \SI{25.6}{kpc}$. The simulations evolve $ 256^3 $ dark-matter particles, each of mass $M_{\mathrm{particle}} = \SI{8.24e8}{{M}_{\odot}}$, in a box of comoving size $L = \num{50} \ h^{-1} \si{Mpc}$ from $z=99$ to $z=0$.\footnote{We make use of the Python package \texttt{pynbody} \citep{pynbody} to analyse the information contained in the simulation snapshots.}

The haloes were identified using the \texttt{SUBFIND} halo finder \citep{gadget}, a friends-of-friends method with a linking length of $0.2$, with the additional requirement that particles in a halo be gravitationally bound. While \texttt{SUBFIND} also identifies substructure within halos, we consider the entire set of bound particles to make up a halo and do not subdivide them further. The simulation contains $18,801$ haloes at $z=0$, ranging from masses of $\sim 10^{9}~\mathrm{M}_{\odot}$ to $ \sim 10^{14}~\mathrm{M}_{\odot}$. 

We used the the final snapshot ($z=0$) to label each particle with its corresponding class. At $z=0$, we split the dark matter particles between two classes; \emph{IN haloes} and \emph{OUT haloes}. We chose the IN class to contain all particles in haloes of mass $M \geq 1.8 \times 10^{12}~\mathrm{M}_{\odot}$ at $z=0$ ($401$ haloes), and the OUT class to contain all remaining particles, including those in haloes of mass $M < 1.8 \times 10^{12}~\mathrm{M}_{\odot}$ and those that do not belong to any halo.\footnote{The mass scale $M=1.8 \times 10^{12}~\mathrm{M}_{\odot}$ corresponds to the mass of a particular halo of the simulation and was chosen as the class boundary for convenience.} This choice was made in order to split the haloes into the two classes at an intermediate scale within the mass range probed by the simulation. Our pipeline allows the selection of any mass threshold which would ultimately allow us to extend the binary classification to a multi-class one.

Each particle, with its associated class label, was traced back to the initial conditions ($z=99$) where we extracted features to be used as input for the random forest as described below. The random forest was trained based on these input features and the known output class for a training subset of particles. We tested the algorithm using the remaining dark matter particles, where the random forest's class prediction was compared to their respective true class label. The robustness of the algorithm was tested further on independent N-body simulations (Sec. \ref{sec:blind}).

\subsection{Density Field Features}
\label{sec:densityfeatures}
Most machine learning algorithms, including random forests, require a \emph{feature extraction} process to extract key properties of the dark matter particles. The classification performance crucially depends on whether or not the chosen features provide meaningful information to allow for a clean separation between the IN and OUT classes. 

We extracted machine learning features from the linear density field. This choice was motivated by the work of \citet{PS} (PS) who developed a model to predict the (comoving) number density of dark-matter haloes as a function of mass based on properties of the linear density field. The ansatz is that a Lagrangian patch will collapse to form a halo of mass $M$ at redshift $z$ if its linear density contrast exceeds a critical value $\delta_c(z)$. An improved theoretical footing for PS theory was developed by \citet{Bond} based on the excursion-set formalism, known as extended Press-Schechter (EPS). The crucial assumption is that the final halo mass corresponds to the matter enclosed in the \textit{largest} possible spherical region with density contrast $\delta_L=\delta_c$. This method yields a halo mass function qualitatively consistent with numerical simulations, suggesting that a useful mapping between Lagrangian regions and final collapsed haloes can be obtained from spherical overdensities. This motivates our choice of machine learning features from the initial linear density field as follows. 

We smoothed the density contrast $ \delta (\textbf{x}) = \left[ \rho (\textbf{x}) - \bar{\rho} \right]/ \bar{\rho} $, where $ \bar{\rho} $ is the mean matter density of the universe, on a smoothing scale $R$,
\begin{equation}
	\delta (\textbf{x}; R) = \int \delta \left( \textbf{x}^\prime \right) W_{\mathrm{TH}} \left( \textbf{x} - \textbf{x}^\prime; R \right) \text{d}^3 x^\prime,
	\label{smoothed_delta}
\end{equation}
where $W_{\mathrm{TH}} (\textbf{x}, R)$ is a real space top-hat window function 
\begin{equation}
	W_{\mathrm{TH}} (\textbf{x},R) = \begin{cases}  
	\dfrac{3}{4 \pi R^3} &\text{ for } \left| \textbf{x} \right| \leq R, \\ 
	0  &\text{ for }  \left| \textbf{x} \right| >R.
	\end{cases}
\end{equation}

The convolution \eqref{smoothed_delta} was carried out in Fourier space, which naturally accounts for the periodicity of simulations. A window function $W(\textbf{x}, R)$ of characteristic radius $R$ corresponds to a mass scale $M_{\mathrm{smoothing}} = \bar{\rho} V(R)$, where in the case of a top-hat window function $V_{\mathrm{TH}}(R) = 4/3 \pi R^3$. The feature for machine learning then consists of the density contrast smoothed with a top-hat window function of mass scale $M_{\mathrm{smoothing}}$ (or, smoothing scale $R$) centred on the particle's position in the initial conditions.

\begin{figure}
	\includegraphics[width=\columnwidth]{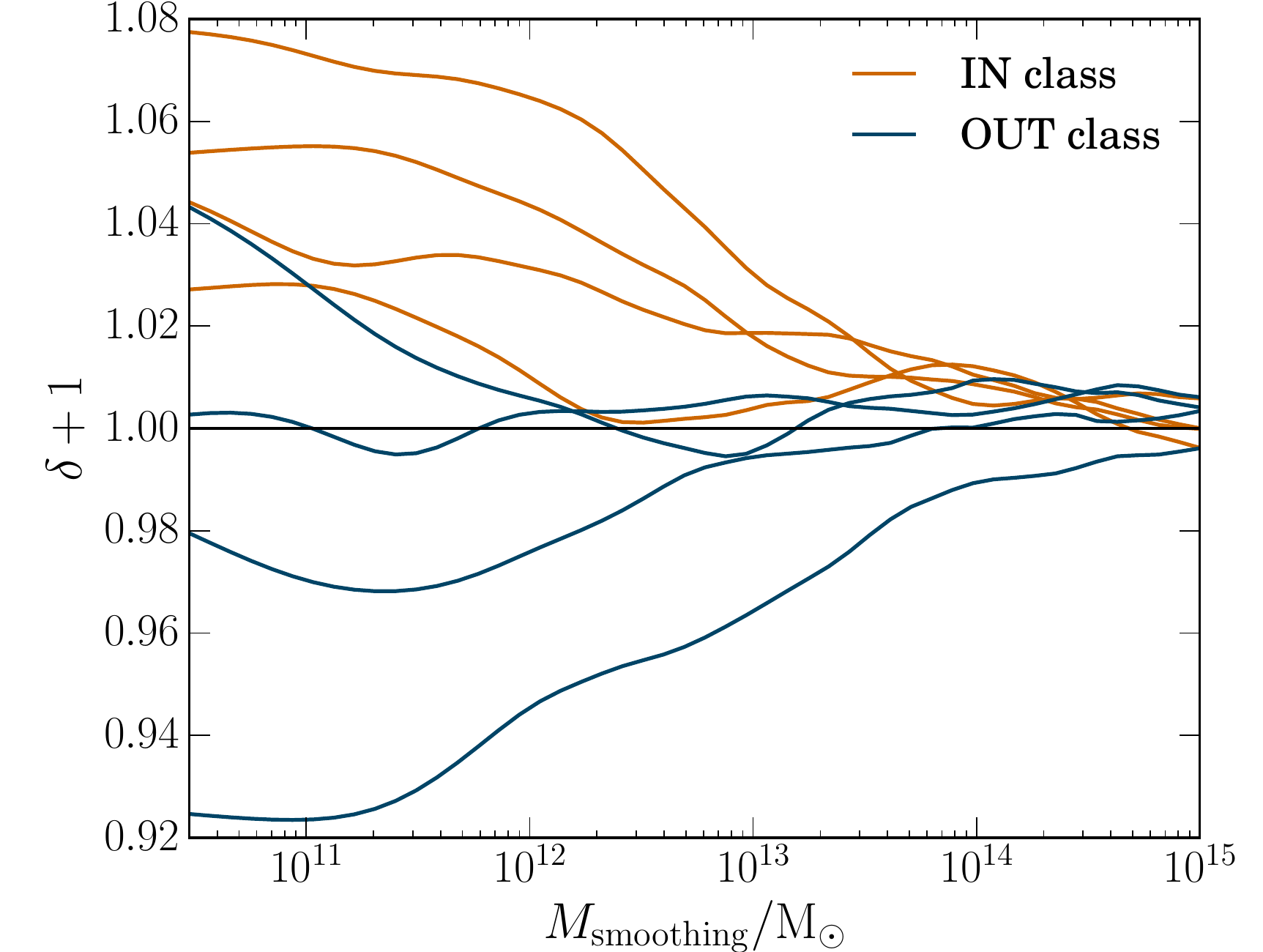}
    \caption{Examples of density trajectories corrresponding to particles belonging to the IN and OUT classes. The linear density field is smoothed with a real space top-hat filter centred on each particle's initial position. We calculate the smoothed overdensity $\delta$ as the smoothing mass scale $M$ is increased.}
    \label{fig:trajectories}
\end{figure}

We repeated the smoothing for $50$ mass scales evenly spaced in $\log M$ within the range allowed by the volume and resolution of the simulation box i.e., $\num{3e10} \leq M_\mathrm{smoothing} / \mathrm{M}_{\odot} \leq \num{1e15}$, yielding a set of $50$ features per particle. We found that using a larger number of smoothing scales did not yield improvement in the classification performance, meaning that $50$ smoothing scales were sufficient to capture the relevant information carried by the density field.

In the context of excursion set theory, the density contrast of a particle as a function of smoothing scale is known as a \textit{density trajectory}. Fig.  \ref{fig:trajectories} shows examples of density trajectories of particles belonging to the true IN and OUT classes. The trajectories describe whether particles are found in overdense or underdense regions as a function of increasing mass scale. As one approaches the largest mass scales probed by the simulation box, the trajectories start to converge to $\delta(x, \infty)=0$, where the density coincides with the mean density of the Universe. The ensemble of trajectories constitutes the full feature set we used to first train then test the random forest.

\subsection{Training the random forest}
\label{sec:RF}
We make use of the random forest implementation in the \textsc{scikit-learn} \citep{sk-learn} Python package. The random forest was trained using a set of  $50$,$000$ randomly selected particles from the simulation, each carrying its own set of density features and corresponding IN or OUT class label. The size of the training set was chosen to form a subset of particles representative of the full simulation box. To test for representativeness, we checked the performance of the algorithm for training sets of different sizes and found no improvement for training sets larger than $50$,$000$ particles. Therefore, we concluded that $50$,$000$ randomly selected particles are sufficient to form a training set representative of the full simulation box. The remaining particles in the simulation were used as a test set; the trained random forest predicts the class label of the particles in the test set, which is then compared to the particles' true labels to assess the algorithm's performance. Note also that random forests are robust to correlated features \citep{Leo}, meaning that the high correlation present in our density features does not affect the predictive performance of the algorithm.

Like most machine learning algorithms, random forests have hyperparameters which need to be optimised for a given training set. These include the number of trees and the maximum depth of the forest, the maximum number of particles at the end node of a tree and the size of the subset of features to select at a node split. We used a grid search algorithm combined with $k$-fold cross validation \citep{kfold} to optimise the random forest's hyperparameters. In $k$-fold cross validation, the training set is divided into $k$ equally sized sets where $k-1$ sets are used for training and one is used as a validation set, on which the algorithm is tested. This procedure is repeated $k$ times so that each set is used as a validation set once. For each validation set we evaluate a score based on a chosen scoring metric (here we use the area under the Receiver Operating Characteristic curve, see Sec. \ref{sec:denclass}) and average scores over all $k$ validation sets to obtain the final score of a training set. Here, we performed a five-fold cross validation for all combinations of hyperparameters and retained the combination which achieved the best score.

\section{Interpreting the classification output}
\label{sec:denclass}

\begin{table}
	\renewcommand{\arraystretch}{1.4}
	\centering
	\caption{Confusion matrix for two classes: Positives and Negatives. We use this to quantify the performance of the machine learning algorithm, where the positives are particles of the IN class and the negatives are particles of the OUT class.}
	\label{tab:confusion}
		\begin{tabular}{|cc|cc|}
			\hline
			& & \multicolumn{2}{c|}{\textbf{True Class}} \\
			& & \textbf{P} & \textbf{N} \\
			\hline
			\multirow{2}{*}{\makecell{\textbf{Predicted} \\ \textbf{Class}}}& \multicolumn{1}{c|}{\textbf{P}} &  
			\multicolumn{1}{c|}{True Positive (TP)} & False Positive (FP) \\
			\cline{3-4}
			& \textbf{N} & \multicolumn{1}{c|}{False Negative (FN)} & True Negative (TN) \\
			\hline
		\end{tabular}
	\renewcommand{\arraystretch}{1}
\end{table}

\begin{figure}
	\includegraphics[width=\columnwidth]{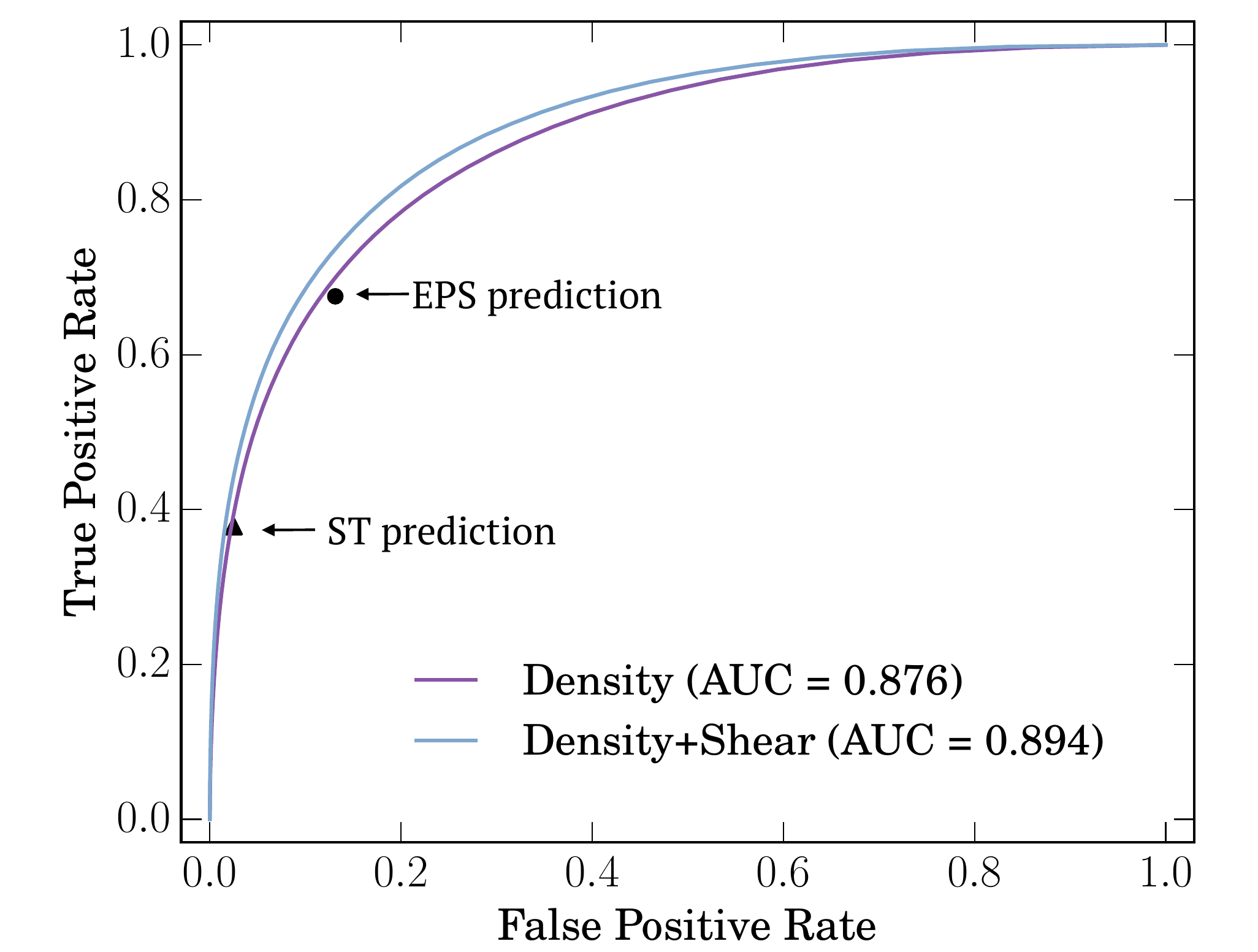}
    \caption{ROC curves for the density feature set and the combined shear and density feature set. The machine learning algorithm is able to learn the information contained in the density trajectories to match the EPS prediction. The ST prediction represents an extension of standard excursion set developed by \citet{ShethTormen}, which adopts a moving collapse barrier motivated by tidal shear effects. The comparison between the two ROC curves shows little improvement in the test set classification once information on the shear field is added. The ST analytic prediction also does not provide an overall improvement compared to the EPS prediction; the false positive rate (or, contamination) decreases at the expense of decreasing the true positive rate (or, completeness). The machine learning algorithm is able to recover the ST analytic prediction when presented with information on the density field alone by altering the probability threshold.}
    \label{fig:ROC_shear_density}
\end{figure}

\begin{figure*}
	\includegraphics[width=0.8\textwidth]{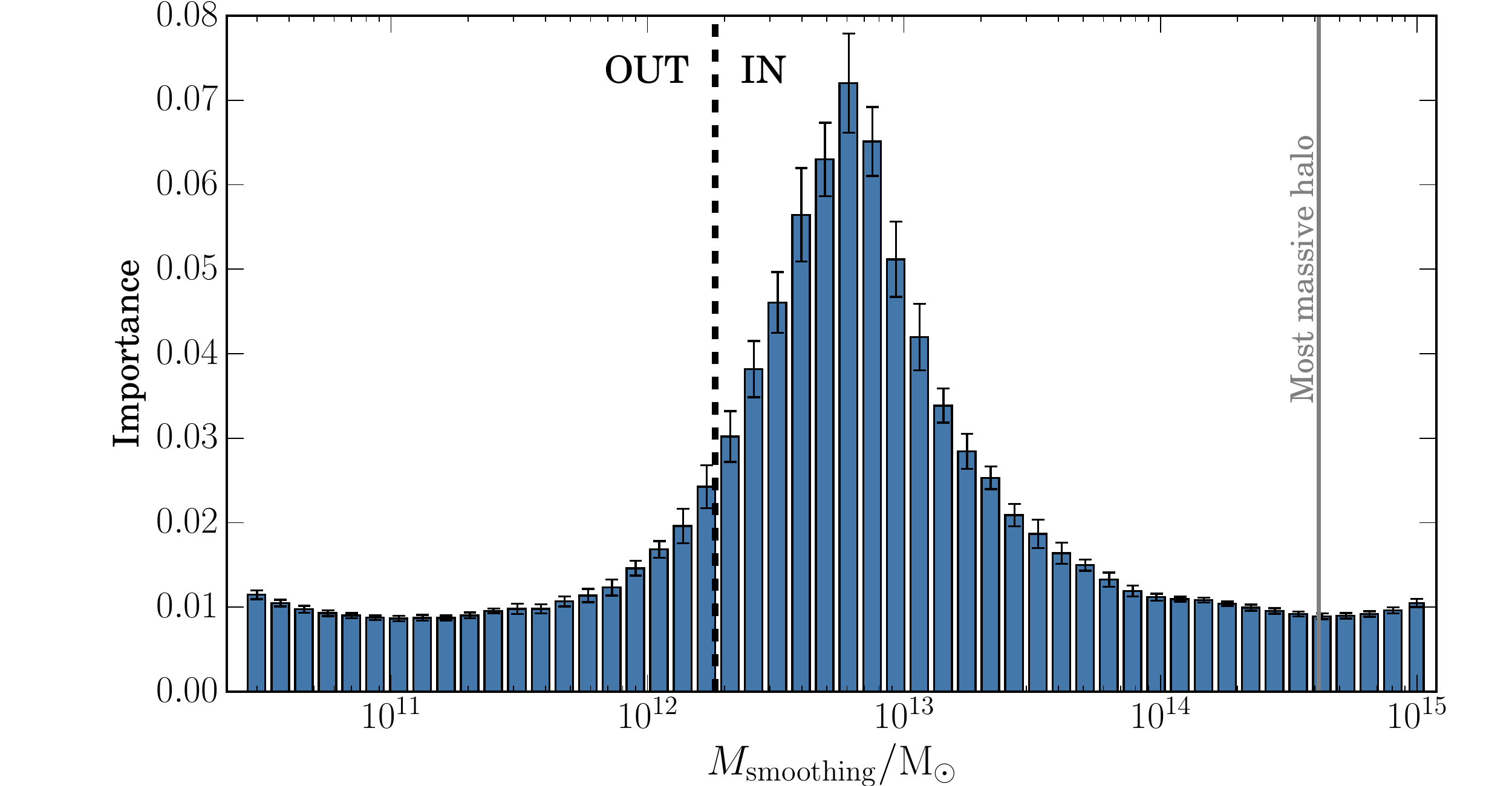}
    \caption{The importance ranking of the density features, shown as a function of their smoothing mass scales. The most relevant information in the training of the random forest comes from the density contrast smoothed at mass scales $10^{12}$ -- $10^{13}$ M$_{\odot}$  scales, within the mass range of the IN class haloes. The largest halo mass in the simulation is marked by a grey line.}
    \label{fig:density_importances}
\end{figure*}

A random forest (like most machine learning algorithms) outputs a probabilistic measure of belonging to a class for every particle. For practical use this must be mapped onto a concrete class for each particle. Many approaches exist for such a mapping; we choose to consider different probability thresholds at which a particle is considered to belong to a class. A high probability threshold will contain a very pure sample of particles but also will be incomplete. As the probability threshold decreases, one allows for a more complete set of particles at the expense of including misclassified ones.

Once the probability-to-class mapping is established, we quantify the performance of the algorithm making use of a confusion matrix for binary classification problems as shown in Table  \ref{tab:confusion}. Throughout this analysis we always take the positives to be particles of the IN class and negatives to be particles of the OUT class. The perfect classifier consists of true positives and true negatives only. A more realistic classifier will include a number of incorrectly classified particles: misclassified positives fall in the false negative category, yielding a loss of \emph{completeness}, and misclassified negatives fall in the false positive category, yielding an increase in \emph{contamination}. 
We measure the true positive rate (TPR), the ratio between the number of particles correctly classified as positives and the total number of positives in the data set, 
\begin{equation}
\mathrm{TPR} = \frac{\mathrm{TP}}{\mathrm{TP} + \mathrm{FN}},
\end{equation}
and the false positive rate (FPR), the ratio between the number of particles incorrectly classified as positives and the total number of negatives in the data set,
\begin{equation}
\mathrm{FPR} = \frac{\mathrm{FP}}{\mathrm{FP} + \mathrm{TN}}.
\end{equation}

Receiver Operating Characteristic (ROC) curves \citep{green1988signal, AUC, Fawcett} are a tool to graphically represent the balance between completeness and contamination at various probability thresholds. A ROC curve compares the true positive rate to the false positive rate as a function of decreasing probability threshold. As one lowers the probability threshold, one allows for a more complete set of IN particles (increase in true positive rate) at the expense of a larger contamination of misclassified particles (increase in false positive rate). The area under the curve (AUC) of a ROC curve is a useful quantity to compare classifiers. The perfect classifier would have an AUC of $1$, whereas a random assignment of classes would obtain an AUC of $0.5$. Typically, algorithms are considered to be performing well if AUC $\geq 0.8$.

We use ROC curves and AUCs to evaluate and compare the performance of the random forest for different feature sets (Sec. \ref{sec:den_results} \& \ref{sec:shear}), different halo mass and radial position ranges (Sec. \ref{sec:part_properties}) and different simulations (Sec. \ref{sec:blind}).

\section{Density field Classification}
\label{sec:den_results}

Figure \ref{fig:ROC_shear_density} shows the ROC curve for the density feature set resulting from classifying all particles in the simulation that were not used for training the random forest. The random forest achieves an AUC score of $0.876$.

In order to assess whether machine learning can learn as much as human-constructed models, we wish to compare its performance to existing theories. In particular, the EPS formalism motivated our choice of density features and has been demonstrated to infer approximately correct number densities of collapsed haloes from a Gaussian random field \citep{Bond}. Although EPS is commonly used to predict the dark matter halo mass function, we make use of it to predict an independent set of class labels for the test set particles and compare their accuracy to that of the machine learning predictions. 

Following EPS, the fraction of haloes of mass $M$ is equivalent to the fraction of density trajectories with a first upcrossing of the density threshold barrier $\delta_{\mathrm{th}}$ at mass scale $M$. We take the density threshold to be the spherical collapse threshold adopted by \citet{Bond}: $ \delta_{\mathrm{th}}(z) = \left( D(z)/D(0) \right) \delta_{\mathrm{sc}} $, where $\delta_{\mathrm{sc}} \approx 1.686 $. The predicted halo mass of each particle is given by the smoothing mass scale of the particle's first upcrossing. We then assign to each particle an IN or OUT label depending on whether its predicted halo mass falls in the mass range of the IN or OUT class. We emphasise that the labels inferred from the EPS framework are independent from the predictions of the random forest.

We plot in Fig. \ref{fig:ROC_shear_density} the resulting true positive rate and false positive rate inferred from the EPS predicted labels and find that the EPS prediction lies on the ROC curve of the random forest. In other words, the random forest is able to `learn' EPS and the EPS results correspond to a $\sim 42\%$ probability threshold on the ROC curve. Machine learning adds the flexibility to trade contamination for completeness along the ROC curve as we vary the probability threshold. Instead, EPS results in a single point in true positive rate-false positive rate space since it gives a single prediction for each particle rather than a probability associated with a class.

\subsection{Physical Interpretation}

The algorithm's performance depends on whether or not the input features contain relevant information to separate particles between classes. For example, the ideal feature would split a set of particles into two pure sets, each containing only particles of one class. By contrast, irrelevant features are not able to distinguish between classes, yielding a poor class separation in the two resulting sets. Therefore, we can determine which features contain the most information in mapping particles into the correct halo mass range, based on their ability to separate classes when training the random forest. 

There are many metrics designed to measure the relevance of the inputs to a machine learning algorithm; here we use \emph{feature importances} \citep{f-imp}. The importance of a feature $X$ is a weighted sum of the impurity decrease\footnote{We use Shannon entropy to measure the impurity at a node $i_E(t) = - \sum\limits_{i=1}^c p(j,t) \log_2 p(j,t)$, where $p(j,t)$ is the proportion of particles that belong to class $j$ at node $t$ and $c$ is the total number of classes.} at all nodes $t$ where the feature is used, averaged over all trees $T$ in the forest:
\begin{equation}
	\mathrm{Imp(X)} = \frac{1}{N_T}\sum_{T} \sum_{t \in T} p(t) \Delta i (t),
\end{equation}
where $N_T$ is the number of trees, $p(t)$ is the fraction of particles reaching node $t$ and $\Delta i(t)$ is the impurity decrease, i.e. the difference in entropy between the parent node and the child nodes.

We calculate the relative importances in the density feature set to find the most relevant features in distinguishing between the IN and OUT classes.  Fig. \ref{fig:density_importances} shows the relative importance of each density feature as a function of its smoothing mass scale. The importances are normalised such that the sum of all importances is $1$ and the errors are computed by training the random forest multiple times, each with a randomly drawn set of training particles. The largest halo mass in the simulation is marked by a grey line. We find that most of the information lies in mass ranges of $10^{12}$ -- $10^{13}~\mathrm{M}_{\odot}$, just above the boundary between the IN and OUT classes.

\begin{figure*}
	\includegraphics[width=0.8\textwidth]{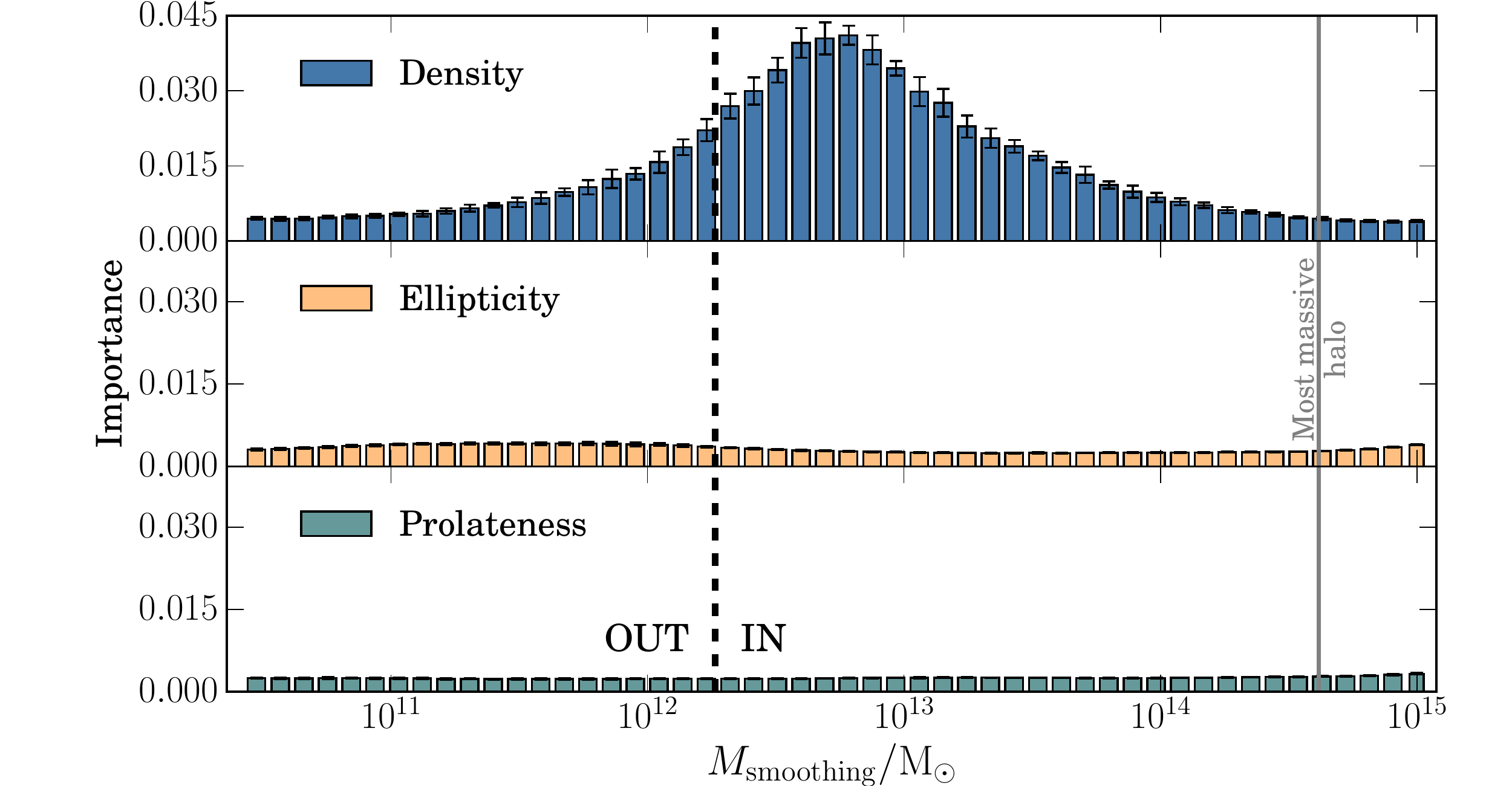}
    \caption{Relative importance of the density features (\textit{upper panel}), ellipticity features (\textit{middle panel}) and prolateness features (\textit{lower panel}) in the full shear and density feature set. The density features are more relevant than the ellipticity and prolateness features. This confirms that the shear field adds little information in distinguishing whether particles will collapse in haloes of mass above the class boundary mass scale or not, compared with the density field.}
    \label{fig:importances_shear_density}
\end{figure*}

\section{Adding the tidal shear tensor}
\label{sec:shear}

Peaks in Gaussian random fields are inherently triaxial \citep{Doroshkevich, BBKS}. Therefore, extensions of the standard spherical model were made in order to incorporate the dynamics of ellipsoidal collapse. The impact of the tidal shear on properties of collapsed regions has been extensively studied \citep{Bond&Myers, ShethTormen, Sheth}. \citet{ShethTormen} (ST) have studied how ellipsoidal collapse modifies the mass function of dark matter haloes in the excursion set formalism. Spheres are distorted into an ellipsoid due to tidal shear effects and the collapse time of a halo therefore depends explicitly on the ellipticity and prolateness of the tidal shear field. 

We extended the original density feature set to incorporate additional information on the local tidal shear field around particles. We studied the impact on the halo classification performance and quantified the shear's relevance in the training process via the feature importances. The advantage of studying tidal shear effects with machine learning is that these can be straightforwardly translated into features and used as input to the same machine learning algorithm. On the other hand, analytic models usually require incorporating approximations to the tidal shear within the excursion set formalism. In general, any potentially relevant physical property can be added in the form of a feature without adding complexity to the algorithm.

We will first describe how we constructed features from the tidal shear field, then present the classification results of the full density and shear feature sets. 

\subsection{Tidal shear features}

The deformation tensor is given by the Hessian of the gravitational potential
\begin{equation}
	D_{ij} = \dfrac{\partial^2 \Phi}{\partial x_i \partial x_j},
	\label{eq:shear}
\end{equation}
where $\Phi(\textbf{x})$ is the peculiar gravitational potential at position $\textbf{x}$ and is related to the density contrast via Poisson's equation $\nabla^2 \Phi = \delta$.

The ordered eigenvalues of $D_{ij}$, $\lambda_1 \geq \lambda_2 \geq \lambda_3$, can be re-parametrised in terms of the ellipticity, $e$, and prolateness, $p$ \citep{Bond&Myers}:
\begin{align}
	e &= \dfrac{\lambda_1 - \lambda_3}{2 \delta}, \\
	p &= \dfrac{\lambda_1 - 2 \lambda_2 + \lambda_3}{2  \delta},
	\label{eq:ell_prol}
\end{align}
where $\lambda_1 + \lambda_2 + \lambda_3 = \delta$ and $\delta$ is the smoothed overdensity used as a density feature.
In order to minimise redundancy between the features, we removed the density dependence from the ellipticity and prolateness. We computed the eigenvalues of the traceless deformation tensor, known as the tidal shear tensor, $t_i = \lambda_i - \delta/3$, now satisfying $t_1 + t_2 + t_3 = 0$. The ellipticity and prolateness in terms of the traceless eigenvalues $t_i$ take the form
\begin{align}
	e_t &= t_1 - t_3, \\
	p_t &= 3 \left( t_1 + t_3 \right).
	\label{eq:ell_prol_features}
\end{align}

For each particle we assigned two new features $e_t$ and $p_t$ evaluated at each smoothing mass scale. Therefore, the original $50$--dimensional feature set of density contrasts was augmented to a $150$--dimensional feature set given by the density contrast, ellipticity and prolateness. To test the robustness of random forests to a high-dimensional feature space, we used PCA to reduce the $150$--dimensional feature set to a $10$--dimensional space retaining $98\%$ of the information contained in the original feature set. We found identical predictive performance, meaning that random forests are robust to a $150$--dimensional feature set.

\subsection{Results}

The ROC curve of the density and shear feature set is overplotted in Fig. \ref{fig:ROC_shear_density}. We find that adding information on the tidal shear tensor shows little improvement compared to the case of the density-only feature set. We find an improvement of only $2\%$ in the AUC of the ROC curve. 
Fig. \ref{fig:importances_shear_density} demonstrates the low impact of the shear features in the classification process. The three panels show the relative importance in the training process of the random forest of the density, ellipticity and prolateness features as a function of smoothing mass scales. The most relevant features are the density contrasts smoothed on mass scales in the range $10^{12}$ -- $10^{13}$ M$_{\odot}$, similar to what was found in the case of the density-only feature set (Fig. \ref{fig:density_importances}). The distributions of the density importances in the two feature sets are consistent despite minor variations in the peak and variance of the distributions. The changes are due to the change in the range of hyperparameters when increasing the dimensionality of the feature set from $50$ to $150$ features. The ellipticity and prolateness have low feature importance scores confirming that the information they contain is irrelevant to the training process of the machine learning algorithm compared with that of the density field. 

As with the density feature set, we can compare the machine learning predictions to existing analytic predictions based on the same set of properties of the initial conditions. The ST formalism provides a prescription to predict the final halo mass of a particle based on the density field and the shear field, which we can use to compare to the machine learning output. 

ST accounts for the effect of the shear field in the context of the excursion set formalism by adopting a moving collapse barrier rather than the spherical collapse barrier adopted by \citet{Bond}. The ST collapse barrier $b(z)$ varies as a function of the mass variance $\sigma^2(M)$ and is given by
\begin{equation}
	b(z) = \sqrt{a} \delta_\mathrm{sc}(z) \left[ 1 + \left( \beta \dfrac{\sigma^2 (M)}{a \delta_{\mathrm{sc}}^2 (z)} \right)^{\gamma} \right], 
	\label{eq:ST_barrier}	
\end{equation}
where $\delta_{\mathrm{sc}} (0) \approx 1.686$, the parameters $\beta  = 0.485$ and $\gamma = 0.615$ incorporate an approximation to ellipsoidal dynamics, and $a = 0.707$ is a normalisation constant. These values are the best-fit parameters found in \citet{Sheth}. The predicted halo mass of each particle follows the excursion-set framework as for the EPS case; the largest mass scale at which the particle's trajectory up-crosses the collapse barrier in Eq. \eqref{eq:ST_barrier} gives the predicted halo mass.

The triangle labelled ``ST prediction'' in Fig. \ref{fig:ROC_shear_density} shows the true and false positive rates predicted by ST. In our study, the ST formalism does not yield an absolute improvement to EPS theory; the false positive rate decreases at the expense of a decrease in the true positive rate. Therefore ST predicts a less contaminated but more incomplete set of IN class particles compared to EPS, corresponding to a probability threshold of $73\%$ on the ROC curve. We find that the random forest is able to reproduce the ST result with both the density-only feature set and the shear and density feature set. This shows that there is sufficient information in the density field for the random forest to match the analytic ST prediction. 

Overall, we find that shear effects do not contain additional physical information to improve the classification output of the random forest. The learning process of the algorithm is predominantly driven by the local overdensity around dark matter particles and unaffected by the surrounding tidal shear. The analytic ST prediction, interpreted as an improvement to standard EPS due to the inclusion of tidal shear effects, can be reproduced by the random forest when trained on the density field only. In conclusion, these results show that the physical processes leading to dark matter halo formation for our choice of mass scale splitting the two classes are insensitive to tidal shear effects in the initial conditions.

\begin{figure*}
	\includegraphics[width=\textwidth]{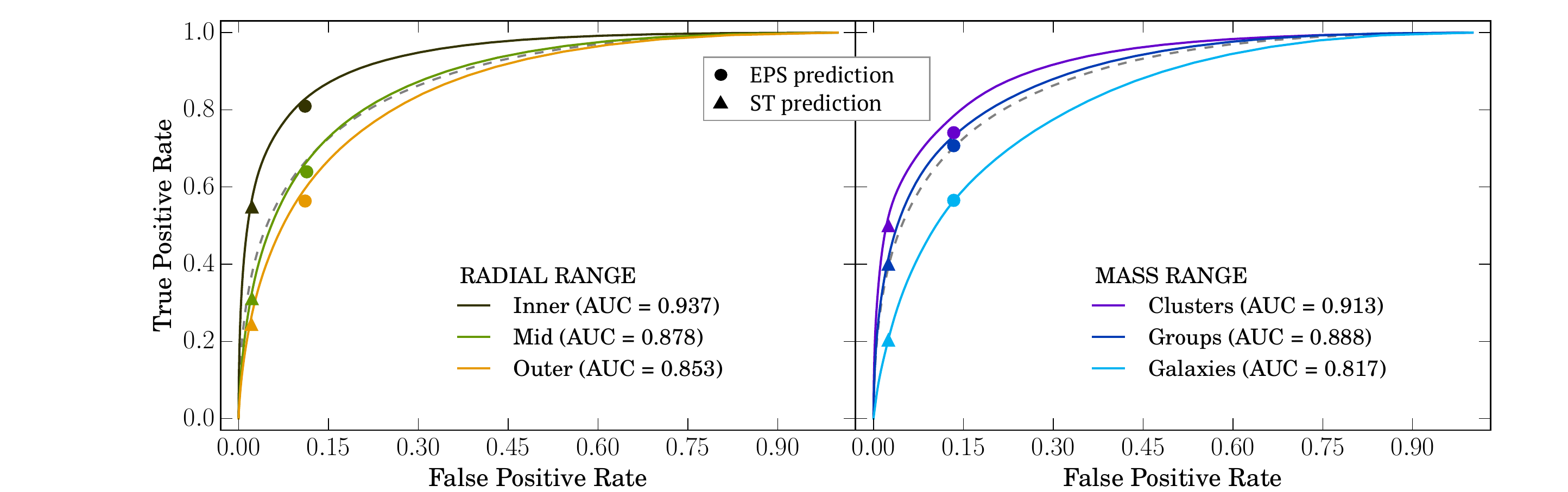}
    \caption{\textit{Left panel}: The IN class particles are split into inner ($ r/r_{\mathrm{vir}} \leq 0.3$), mid ($0.3 < r/r_{\mathrm{vir}} \leq 0.6$) and outer ($0.6 < r/r_{\mathrm{vir}} \leq 1$) radial ranges according to their distance from the centre of the halo. The ROC curves for each category show that the classification performance improves for particles closer to the halo's centre of mass. \textit{Right panel}: The IN class particles are split into cluster-sized ($\num{1e14} \leq M_{\mathrm{halo}} /\mathrm{M}_{\odot} \leq \num{4e14}$), group-sized ($\num{1e13} \leq M_{\mathrm{halo}} /\mathrm{M}_{\odot} < \num{1e14}$) and galaxy-sized ($\num{1.2e12} \leq M_{\mathrm{halo}} /\mathrm{M}_{\odot} < \num{1e13}$) haloes and the ROC curves show the random forest's performance in classifying each category. Particles in higher mass haloes are increasingly better classified by the random forest. The ROC curve of the full test set of particles is shown as a dashed line in both panels for comparison. The EPS and ST predictions, labelled by dots and triangles respectively, are also overplotted for each halo mass and radial position category.}
    \label{fig:ROC_mass_radius_bins}
\end{figure*}

\section{Classification dependence on halo mass and radial position}
\label{sec:part_properties}
We now investigate how properties of particles such as the position within a halo and the halo mass affect the accuracy of classification when the algorithm is trained on density features only. To do this we split the test particles into categories based on their radial and halo mass properties to study their respective classification performance.

First, we subdivided particles of the IN class into three mass ranges: particles in \textit{cluster}-sized haloes ($\num{1e14} \leq M_{\mathrm{halo}} /\mathrm{M}_{\odot} \leq \num{4e14}$), particles in \textit{group}-sized haloes ($\num{1e13} \leq M_{\mathrm{halo}} /\mathrm{M}_{\odot} < \num{1e14}$) and particles in \textit{galaxy}-sized haloes ($\num{1.2e12} \leq M_{\mathrm{halo}} /\mathrm{M}_{\odot} < \num{1e13}$). We combined each of these subsets in turn with all the OUT particles to form three distinct test sets.

The ROC curves for the three mass range categories of haloes are shown in the right panel of Fig. \ref{fig:ROC_mass_radius_bins}, where the ROC curve of the full original test set is shown for comparison (dashed line). We find that particles in cluster-sized haloes reach an AUC of $0.913$, whilst particles in group-sized haloes and galaxy-sized haloes are increasingly more difficult to classify. We overplotted the ST (triangles) and EPS (dots) predictions for each halo mass category of particles, again showing results consistent with those of the machine learning algorithm.

\begin{figure}
	\includegraphics[width=\columnwidth]{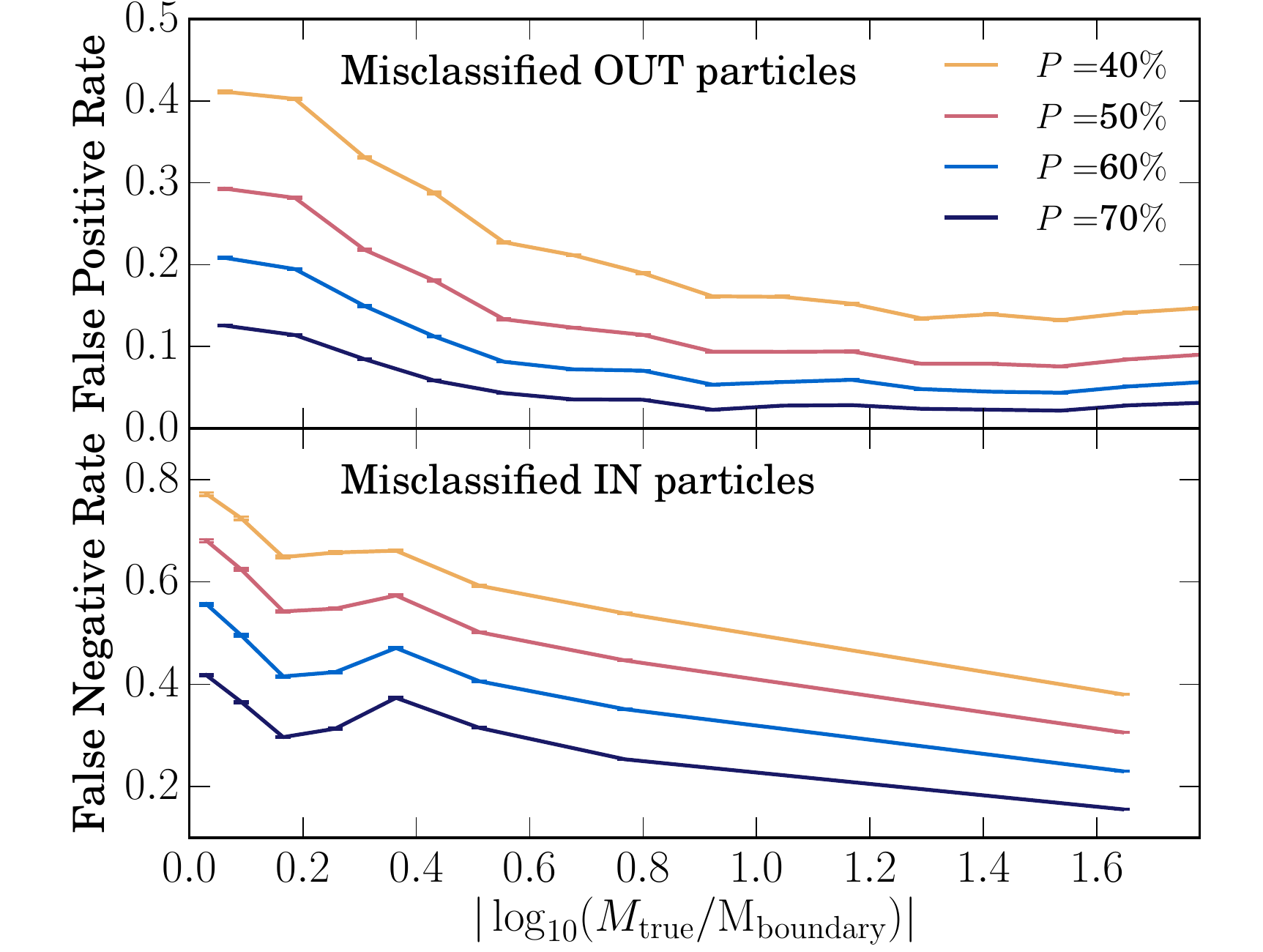}
    \caption{Fraction of misclassified particles in haloes of each mass bin range, where the halo mass bins are labelled as a function of their distance from the IN/OUT boundary mass scale. The upper (lower) panel shows the fraction of misclassified OUT (IN) particles i.e., the false positive (negative) rate in each mass bin. We consider four distinct probability thresholds for assigning a particle's (IN or OUT) class, where higher thresholds imply lower contamination. The misclassification rate increases as the true mass approaches the classification boundary for all choices of the completeness-to-contamination trade-off.}
    \label{fig:hm_dependence}
\end{figure}

It is likely that the decrease in performance as a function of halo mass is a result of the choice of mass scale used to split haloes into classes, $M = 1.8 \times 10^{12}~\mathrm{M}_{\odot}$. This was a necessary step in order to define the two classes of the binary classification problem. Haloes of mass just above and below the IN/OUT mass boundary belong to different classes although they originate from Lagrangian regions with similar properties reflecting their similarity in mass. Therefore, the closer haloes of different classes are in mass, the harder it is for the random forest to distinguish whether their particles belong to one class or the other. Fig. \ref{fig:hm_dependence} further demonstrates that haloes of mass approaching the IN/OUT mass boundary from above and below contain a larger fraction of misclassified particles. In the upper (lower) panel, we show the false positive (negative) rate i.e., the ratio of misclassified OUT (IN) particles over all particles contained in each halo mass bin, for $4$ different probability thresholds. The true halo mass of each particle is shown on the horizontal axis in terms of its distance from the IN/OUT mass boundary. We find that the false positive and negative rates increase for particles in haloes of mass approaching the IN/OUT mass boundary.

We next investigated possible correlations between the particles' position within the haloes and the random forest's classification performance. Here, we subdivided particles of the true IN class into three radial ranges, subject to their radial position in the halo with respect to the halo's virial radius $r_{\mathrm{vir}}$. We defined particles in the \emph{inner radial} range ($ r/r_{\mathrm{vir}} \leq 0.3$), particles in the \emph{mid radial} range ($0.3 < r/r_{\mathrm{vir}} \leq 0.6$) and particles in the \emph{outer radial} range ($0.6 < r/r_{\mathrm{vir}} \leq 1$). Similar to the mass range study, each subset of haloes was combined with all the OUT class particles from the original set to form three distinct sets.

The left panel of Fig. \ref{fig:ROC_mass_radius_bins} shows the ROC curves for the three radial categories, together with that of the original test set again shown for comparison (dashed line). Particles in the innermost regions of haloes are the best classified by the random forest, achieving an AUC of $0.937$ which is greater than that obtained when classifying \emph{all} particles in the simulation. The classification performance of the random forest decreases as we move from the halo's centre-of-mass towards the virial radius. 

We first tested whether the decrease in performance when classifying particles of the outer radial range was due to under-representativeness in the training set. Indeed, if the training particles of the outer radial range are not representative of the entire simulation, the classifier's performance on the outer radial range test set would be strongly affected. To test this, we re-trained the machine learning algorithm with a training set containing equal number of particles for each radial range category. We found identical ROC curves and AUCs as in the left panel of Fig. \ref{fig:ROC_mass_radius_bins}, therefore excluding the possibility that the higher misclassification rate of outer radial range particles is due to non-representativeness in the training set.

One other possible reason may be that particles living in outer regions of haloes are more likely to have been affected by late-time halo mergers, tidal stripping or accretion events. Therefore, the final halo mass prediction for such particles is the result of a more complicated dynamical history involving these late-time effects. Conversely, particles near the halo's centre-of-mass are less sensitive to the halo's assembly history and their final halo mass prediction correlates more strongly with the local overdensity in the initial conditions. This hypothesis could be verified by adding features sensitive to the particles' dynamical history (for instance a particle's initial distance to the nearest density peak) and testing whether this information improves the classification of particles located at the boundary of the halo's virial region. In addition to this, the further particles are from the centre of haloes, the closer they are to the boundary between the IN and OUT classes, where particles are harder to classify for the machine learning algorithm. This also translates into a larger uncertainty in the halo mass prediction for particles at the edge of haloes compared to those in the innermost regions of haloes. As a result, the overall uncertainty in the halo mass predictions of centre-of-mass particles is smaller than for particles in the outskirts of haloes. This result is also consistent with excursion set predictions, where ST demonstrated that centre-of-mass particles provide a better estimate of the final halo mass compared to inferences made from the full ensemble of particles in the simulation. 
To confirm this, we overplotted the EPS (dots) and ST (triangles) predictions for the three radial test sets in the left panel of Fig. \ref{fig:ROC_mass_radius_bins}, demonstrating that analytic formalisms also perform increasingly well for particles that are close to the halo's centre-of-mass. The machine learning algorithm again shows its ability to match the excursion set predictions at fixed probability thresholds for each radial range category.

For completeness, we also explored the misclassification rate of OUT particles that do not belong to any halo. We find that overall these particles have very low misclassification rates compared to particles in haloes. For example, if we consider probability thresholds of $70\%$, $60\%$, $50\%$ and $40\%$ to assign particles to the IN class (as in the upper panel of Fig. \ref{fig:hm_dependence}), the fraction of misclassified over all particles that don't belong to haloes is $2.45\%$, $4.3\%$, $6.58\%$ and $10.11\%$, respectively. Therefore, the OUT particles predicted by the random forest form a highly pure and complete set.

In conclusion, we find that the best classified categories of particles are those which are further away from the classification boundary, both in terms of mass and radius: particles in the most massive and least massive haloes in the simulation; particles in the innermost regions of haloes; and those furthest away in voids. We further tested whether the addition of the tidal shear information could improve the classification performance of poorly classified particles, such as those in the outskirts of halos and in galaxy-sized halos. We find no significant improvement in the classification performance of such particles, other than the $2\%$ improvement found for the whole ensemble and reflected in each mass and radial category.

\section{Blind tests on independent simulations}
\label{sec:blind}
Up to this point we have trained and tested the machine learning algorithm on a single dark-matter-only simulation. To test whether the machine learning algorithm trained on one simulation also gives robust results for different N-body simulations without re-training, we performed blind tests of our pipeline on two independent simulations from the one used for training. 

The first independent test simulation (W-Test) is a different realisation of the same WMAP5 $\Lambda$CDM cosmology adopted in the training simulation, for a box of also same size and resolution (see Sec. \ref{sec:method}). The second independent test simulation (P-Test) is a realisation of a different cosmological model, a \textit{Planck} $\Lambda$CDM cosmology\footnote{The cosmological parameters are $\Omega_{\Lambda} = 0.6914$, $\Omega_{\mathrm{m}} = 0.3086$, $\Omega_{\mathrm{b}} = 0.045$, $\sigma_{8} = 0.831$, $h = 0.6727$, $n_s = 0.96$.}\citep{Planck2015} in a box of comoving size $L = \SI{50}{Mpc}$ containing $N=512^3$ particles. Moreover, in the P-Test simulation we identify haloes at $z=0$ using the Amiga Halo Finder (AHF) \citep{AHF2004, AHF2009}, instead of the \texttt{SUBFIND} halo finder used in both the training simulation and the W-Test simulation. This allows us to simultaneously test the sensitivity of the machine learning algorithm to the choice of halo finder. For each test simulation, we extracted the input features from the initial conditions and used the pre-trained machine learning algorithm to predict the class labels of the simulations' dark matter particles.

\begin{figure}
	\includegraphics[width=\columnwidth]{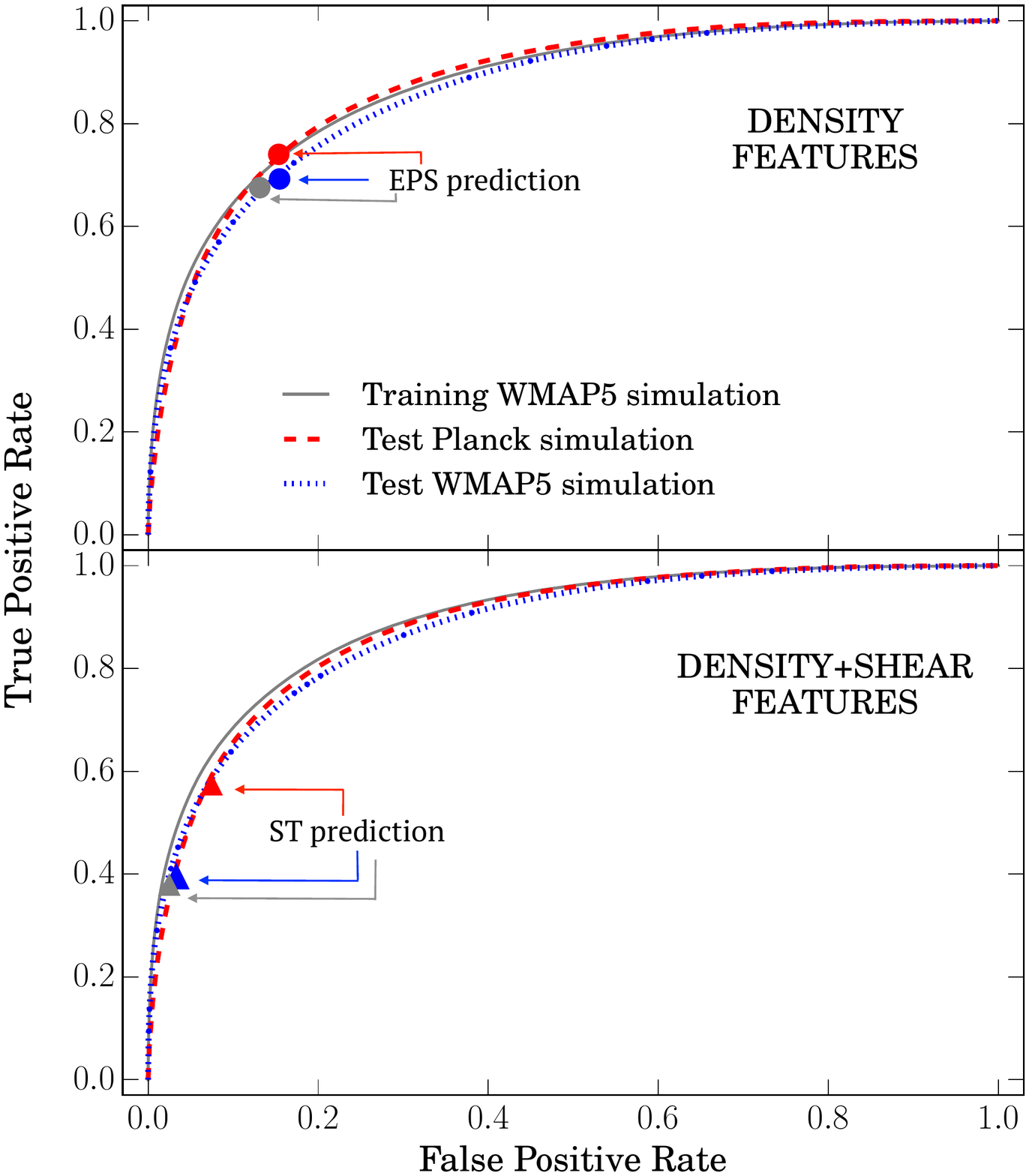}
    \caption{We perform a blind test of the trained machine learning algorithm on two independent N-body simulations; a different realisation of the WMAP5  cosmology used in the training simulation, and a realisation of a \textit{Planck} cosmological model. The ROC curves are consistent in all three simulations for both the density feature set and the density and shear feature set, with differences in the AUCs of order $\sim 1\%$. The EPS and ST predictions in each simulation match the machine learning performance at different probability thresholds, such that the ST formalism always predicts a less contaminated but more incomplete set of IN particles. These blind tests demonstrate the robustness of the results from a machine learning algorithm trained on one simulation, and applied to different realisations of the same cosmology or realisations of different cosmologies.}
    \label{fig:ROC_blind}
\end{figure}

In Fig. \ref{fig:ROC_blind} we compare the performance of the machine learning algorithm for the independent W-Test and P-Test simulations with that of the test set of particles in the training simulation. The upper panel shows the ROC curves obtained from predictions based on the density features only, whilst the lower panel shows the case of density and shear features. The machine learning algorithm produces consistent ROC curves in all three simulations for both feature sets. The P-Test simulation yields a difference in AUC with the training simulation of $0.2\%$  for the density-only feature set and $1.1\%$ for the density and shear feature set. For the W-Test simulation, the AUC difference with the training simulation is of $1.3\%$ for the density-only feature set and $1.6\%$ for the density and shear feature set. Such differences between the test and training simulations are consistent with uncertainties in the AUC due to statistical noise.

The EPS and ST predicted labels are calculated from the first upcrossings of each simulation's respective particles' trajectories. In all three simulations, the machine learning algorithm is able to match the analytic predictions at different probability thresholds, such that the ST formalism consistently predicts a less contaminated but more incomplete set of IN class particles.  For the W-Test simulation, the EPS and ST predictions match the machine learning predictions at probability thresholds of $41.5\%$ and $74.5\%$ respectively, differing only slightly to the $42.8\%$ and $74.7\%$ probability thresholds of the training simulation. For the P-Test simulation, the match to the EPS and ST predictions is found at the lower probability thresholds of $40\%$ and $56\%$, respectively. This is because the change in cosmological parameters in the \textit{Planck} simulation results in a slightly lower EPS collapse barrier and a significantly lower ST collapse barrier compared to those in a WMAP5 cosmological setting. Therefore, trajectories in the P-Test simulation upcross the collapse barriers at larger smoothing mass scales, resulting in more complete but also less pure sets of predicted IN particles. The change in completeness and contamination is such that both the ST and EPS predictions still match the machine learning ROC curves of the P-Test simulation, but for lower probability thresholds than the WMAP5 simulations.

We conclude that the mapping learnt by the algorithm on one simulation can be generalised to different simulations based on the same or different cosmological parameters, without the need for re-training, and that the results are insensitive to simulation settings.

\section{Conclusions}
\label{sec:conclusions}
We have presented a machine learning approach to investigate the physics of dark matter halo formation. We trained the algorithm on N-body simulations, from which it learns to predict whether regions of an initial density field later collapse into haloes of a given mass range. This generated a mapping between the initial conditions and final haloes that would result from non-linear evolution, without the need to adopt halo collapse approximations. Our approach provided new physical insight into halo collapse, in particular in understanding which aspects of the initial linear density field contain relevant information on the formation of dark matter haloes.

We provided the algorithm with a set of properties describing the local environment around dark matter particles. By studying the performance of the algorithm in response to different inputs, insights can be gained into the physics relevant to dark matter halo formation. When the algorithm was trained on spherical overdensities from the linear density field, we found that it matched predictions based on EPS theory. When providing the algorithm with additional information on the tidal shear field (motivated by ellipsoidal collapse approximations), the classification performance of the machine learning was not enhanced. We showed that, for the mass threshold considered in our classification problem, the Sheth-Tormen ellipsoidal collapse model can be recovered from spherical overdensities alone, with predictions that differ from those of EPS theory only in the completeness-to-contamination trade-off. By performing blind analyses of our pipeline, we confirmed the generality of our results for independent initial conditions realisations and variations in cosmological parameters. We conclude that the linear density field contains sufficient information to predict the formation of dark matter haloes at the accuracy of existing spherical and ellipsoidal collapse analytic frameworks.

While the focus of this paper has been on the density field and tidal shear field, any additional property of interest can be extracted from the initial conditions and used as input to the same machine learning algorithm. This allows for straightforward extensions of the present work to investigate the physics of dark matter halo formation further. Future work could also extend the binary classification problem presented in this work into multi-class classification or regression problems. Potential applications of such an extended framework include a new approach to obtaining a halo mass function, which can be directly tested against existing fitting formulae adopted by analytic approaches. More sophisticated machine learning algorithms such as deep learning offer the ability to learn from the training data which features are the most relevant to cosmological structure formation, and future work will investigate their suitability for structure formation studies.

\section*{Acknowledgements}

LLS thanks Nina Roth for providing one of the simulations used in this work and for useful discussions. LLS was supported by the Science and Technology Facilities Council. HVP was partially supported by the European Research Council (ERC) under the European Community's Seventh Framework Programme (FP7/2007-2013)/ERC grant agreement number 306478- CosmicDawn. AP was supported by the Royal Society. ML acknowledges support from the SKA, NRF and AIMS. This work was partially enabled by funding from the UCL Cosmoparticle Initiative.




\bibliographystyle{mnras}
\bibliography{bib_mlhalos} 





\bsp	
\label{lastpage}
\end{document}